\documentclass[sigplan]{acmart}
\usepackage{csquotes}
\AtBeginDocument{%
  \providecommand\BibTeX{{%
    \normalfont B\kern-0.5em{\scshape i\kern-0.25em b}\kern-0.8em\TeX}}}

\setcopyright{acmcopyright}
\copyrightyear{2018}
\acmYear{2018}
\acmDOI{XXXXXXX.XXXXXXX}

\acmConference[Conference acronym 'XX]{Make sure to enter the correct
  conference title from your rights confirmation emai}{June 03--05,
  2018}{Woodstock, NY}
\acmPrice{15.00}
\acmISBN{978-1-4503-XXXX-X/18/06}

\copyrightyear{2022} 
\acmYear{2022} 
\setcopyright{acmcopyright}\acmConference[WMSSH '22]{1st ACM Workshop on Mobile and Wireless Sensing for Smart Healthcare}{October 21, 2022}{Sydney, NSW, Australia}
\acmBooktitle{1st ACM Workshop on Mobile and Wireless Sensing for Smart Healthcare (WMSSH '22), October 21, 2022, Sydney, NSW, Australia}
\acmPrice{15.00}
\acmDOI{10.1145/3556551.3561191}
\acmISBN{978-1-4503-9520-5/22/10}
\begin{document}

\title{AI enabled RPM for Mental Health Facility}

\author{Thanveer Shaik}
\email{Thanveer.Shaik@usq.edu.au}
\orcid{0000-0002-9730-665X}
\affiliation{%
  \institution{University of Southern Queensland}
  \city{Toowoomba}
  \country{Australia}
}

\author{Xiaohui Tao}
\email{Xiaohui.Tao@usq.edu.au}
\affiliation{%
  \institution{University of Southern Queensland}
  \city{Toowoomba}
  \country{Australia}
}

\author{Niall Higgins}
\email{Niall.Higgins@health.qld.gov.au}
\affiliation{%
  \institution{Royal Brisbane and Women’s Hospital}
  \institution{Queensland University of Technology}
  \city{Brisbane}
  \country{Australia}
}

\author{Haoran Xie}
\email{hrxie@ln.edu.hk}
\affiliation{%
  \institution{Lingnan University}
  \city{Tuen Mun}
  \country{Hong Kong}}
  
\author{Raj Gururajan}
\email{Raj.Gururajan@usq.edu.au}
\affiliation{%
  \institution{University of Southern Queensland}
  \city{Springfield}
  \country{Australia}}

\author{Xujuan Zhou}
\email{Xujuan.Zhou@usq.edu.au}
\affiliation{%
  \institution{University of Southern Queensland}
  \city{Springfield}
  \country{Australia}}

\renewcommand{\shortauthors}{Shaik et al.}

\begin{abstract}
Mental healthcare is one of the prominent parts of the healthcare industry with alarming concerns related to patients' depression, stress leading to self-harm and threat to fellow patients and medical staff. To provide a therapeutic environment for both patients and staff, aggressive or agitated patients need to be monitored remotely and track their vital signs and physical activities continuously. Remote patient monitoring (RPM) using non-invasive technology could enable contactless monitoring of acutely ill patients in a mental health facility. Enabling the RPM system with AI unlocks a predictive environment in which future vital signs of the patients can be forecasted. This paper discusses an AI-enabled RPM system framework with a non-invasive digital technology RFID using its in-built NCS mechanism to retrieve vital signs and physical actions of patients. Based on the retrieved time series data, future vital signs of patients for the upcoming 3 hours and classify their physical actions into 10 labelled physical activities. This framework assists to avoid any unforeseen clinical disasters and take precautionary measures with medical intervention at right time. A case study of a middle-aged PTSD patient treated with the AI-enabled RPM system is demonstrated in this study. 
\end{abstract}




\maketitle

\section{Introduction}
One of the most important goals of hospitalised patient psychiatric care for mental illness patients and depressed suicidal tendency patients is to provide a safe and therapeutic environment for both patients and staff. To manage or minimise acute behavioural disturbance, strategies include non-pharmacological approaches, and integral to this is the continuing need to focus on good communication and teamwork~\cite{emmerson2007contemporary}. Remote patient monitoring (RPM) would be part of a non-pharmacological approach to detect the patient's motion and monitor their vital signs like respiratory rate, heart rate, and pulse rate. This would help in preventing incidents like when patient frustration escalates to physical violence to hurt fellow patients or staff or incidents like depression leading to suicides which are self-harm and the overall milieu of the inpatient unit~\cite{higgins2018implementation}. Some aggressive or agitated patients are difficult to sedate, and even multiple doses of sedating anti-psychotics could result in harm to the patient and staff~\cite{knott2008sedation}. This could lead to incomplete sets of vital signs that limit the identification of deteriorating patients~\cite{cardona2016vital}. There is substantial evidence that alterations in respiratory rate can be used to predict potentially serious clinical events~\cite{cretikos2008respiratory}. Hence, continuous patient monitoring technology has the potential to enhance the clinician’s ability to monitor patients and detect deterioration early. This would help society to avoid patients’ self-harm or physical violence against nursing staff or fellow patients.
\begin{figure*}
  \includegraphics[width=\textwidth]{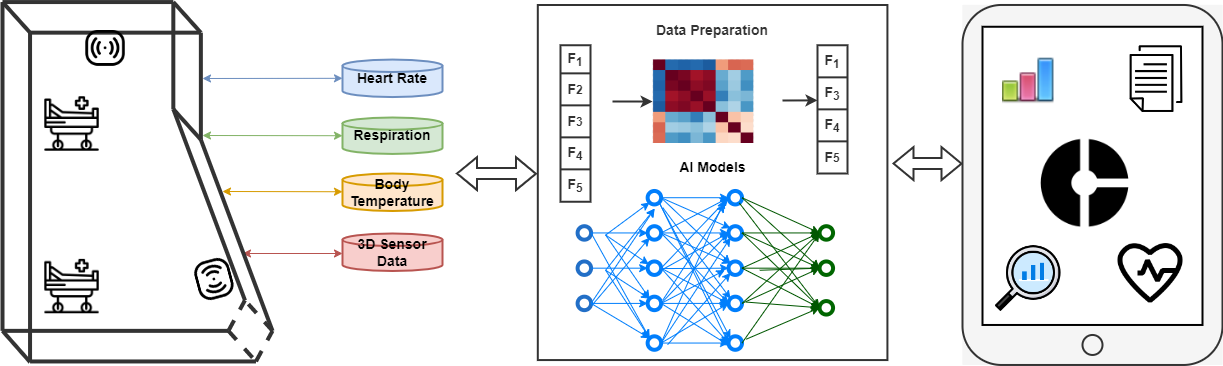}
  \caption{Graphical Abstract}
  \label{fig:teaser}
\end{figure*}
New technological advances in Artificial Intelligence (AI) with modern remote monitoring provide a means of capturing clinical observations of a patient’s health status. RPM enables with AI is gaining more attention with its predictive capability to build an enhanced decision support system for assisting clinicians. Traditional RPM is more intrusive to have dedicated sensors on patients' bodies, which might cause inconvenience to acutely ill patients. Recent innovations like near-field coherent sensing (NCS) have demonstrated the ability in tracking human vital signs using Radio Frequency Identification (RFID)~\cite{bianco2021survey} tags and reader-antennas~\cite{8210832}. This would also assist in elderly health monitoring to monitor their physical actions to avoid any falls~\cite{wickramasinghe2017ambulatory}. The AI-enabled RPM system is being developed to revolutionise patient care via tracking vital signs and detecting physical activities of acutely ill patients in psychiatric care~\cite{Tao2021}. The primary aim of the study is to perform observations of patients' vital signs and physical actions using non-invasive technology, predict future vital signs, and retrieve them to a handheld tablet of the medical staff. 

This paper introduces an end-to-end patient monitoring system configured with AI to collect the vital signs and physical actions of acutely ill patients in a contained mental health facility where patients are admitted to mental health facilities. The proposed framework could assist medical staff in tracking their patients' health status without even touching their bodies by placing passive RFID tags on clothes in different areas of the body. Using AI models, the staff can retrieve their patients' current health status and forecasted vital signs onto their handheld tablets and intervene at the right time to avoid any unanticipated issues related to self-harm or clinical deterioration events as shown in Fig.~\ref{fig:teaser}. The contributions of this study are:
\begin{itemize}
    \item To provide contact-less health monitoring of aggressive or agitated patients in a mental health facility.
    \item A multi-type data processing and information fusion for comprehensive monitoring of patient's health.
    \item To forecast the patients' vital signs for the upcoming 3 hours using prediction modelling.
    \item To monitor the current physical status of patients by classifying their actions.
    
\end{itemize}

The rest of the paper is organised as follows: Section~\ref{relatedworks} presents the related work in early detection of clinical deterioration and existing works of RPM using RFID technology in combination with AI for prediction and classification. Section~\ref{researchproblem} discussed the research problem being focused on in this study. In Section~\ref{framework}, the proposed research framework for an AI-enabled RPM system is discussed in detail along with the prediction and classification models used in this study. A case study adopting the proposed framework is presented in Section~\ref{casestudy}. Finally, this paper is concluded and suggested future work to further enhance the RPM system in Section~\ref{conclusion}.

\section{Related Works}\label{relatedworks}

Patient safety is one of the concerns in global public health. Misidentifying patients in healthcare leads to medical errors in hospitals, causing a major risk to a patient’s safety. To overcome this, advanced tracking technology like RFID  can be implemented to build a wristband for patients using passive RFID tags. Patient details like name, age, blood type, allergies, treatments required, and insurance can be retrieved by scanning their passive tag with an RFID reader~\cite{haddara2018rfid}. Implementing smart identification would assist both patients and medical staff and enhance the safety measures in a hospital~\cite{paaske2017benefits}. However, the limitation of this smart application is wristband contact with patients' skin, and they may resist holding the wristband in aggressive situations. But RFID technology could help in identifying the patients.

Near-field Coherent Sensing (NCS) mechanism in RFID tags has been discovered by researchers at Cornell University. The mechanism is based on electromagnetic energy, in which mechanical motion on the surface and inside a body is modulated onto multiplexed radio signals. This helps to monitor the mechanical motion of the internal organs and reflected RFID tags capture magnetic values. There are limited sensing capabilities and sampling rates in existing systems like electrocardiogram (ECG) and this may compromise monitoring the heart rate, respiration, breath rate, and blood pressure~\cite{hui2018monitoring}. Furthermore, the ECG and acoustics methods might limit the comfort, body motion, and wearing convenience as they require direct contact with patients' skin and effects long-term wearing~\cite{hui2018mitigation}. Sharma et al.~\cite{sharma2018sleep} deployed an RFID passive tag in the chest area to track heart rate, breath rhythm and body motion using the NCS mechanism. In this experiment, the quality of sleep was assessed based on the vital signs' heartbeat, respiration, and upper body motion together. The authors conducted semi-supervised learning to classify the body motion using a support vector machine (SVM) and achieved an accuracy of 91.06 percent.

The research community is working beyond the identification of vulnerable patients to ensure their care and safety. It includes tracking mentally depressed patients with suicidal tendencies. Even in psychiatric care, patients can be managed in a smarter way and provide medical treatment without any delay~\cite{ariffin2015psychiatric}. To monitor patients, data related to patient's health status is required and this can be extracted by integrating RFID technology with the internet of things~\cite{yang2022review}, machine learning and Artificial Intelligence (AI). From RFID applications in health care, tracking and monitoring of patients, are most of the researchers and hospitals explored. Without any medical and human errors, patients can be served with safety backup~\cite{kranzfelder2012real}. Kranzfelder et al.~\cite{kranzfelder2012real} proposed RFID technology for tracking and monitoring retained surgical sponges and surgeons in operating rooms. The authors used passive tags in the experiment for stationary surgical sponges and active tags for the operating team. 

Health monitoring also includes detecting bed and chair exits of hospitalised elderly people to prevent falls. Shinmoto et al.~\cite{shinmoto2017battery,cheung2022night,ganesan2022elderly} proposed a battery-less and wireless wearable sensor system to prevent elderly people fall in the hospital. The proposed battery-less and wireless wearable sensor system were able to achieve a precision of 66.8\%, recall of 81.4\%, and F1-score of 72.4\% for  joint chair and bed recognition. Zhao et al.~\cite{zhao2015emod} built Efficient Motion Detection of Device Free Object (EMoD) to detect and track device-free objects by deploying a few pairs of tags at critical places.

The related works provided a comprehensive understanding of AI strategies implemented in RPM systems. Also, there is a need to adopt these strategies along with RPM systems to monitor mental health patients without intervening in their daily activities. The proposed AI-enabled framework aims to set a benchmark for any subsequent study for enhanced RPM systems.

\begin{figure*}
    \centering
    \includegraphics[width=\textwidth]{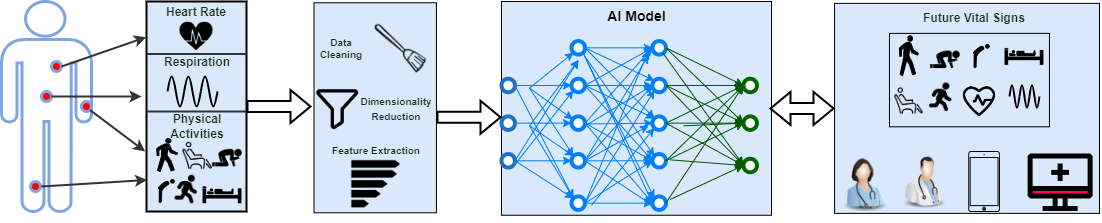}
    \caption{Proposed research framework}
    \label{fig:framework}
\end{figure*}
\section{Research Problem}\label{researchproblem}
The research problem is to track patients' vital signs and monitor their physical actions without hindering their daily activities. Based on past and present time-series data of vital signs, future vital signs will need to be predicted, and physical activities need to be classified. This study is to perform vital signs observation of possibly over-sedated and potentially aggressive patients in a secure, contained mental health facility. Vital signs such as heart rate, respiration, and physical actions will need to be monitored using non-invasive technology. Further to this, build a decision support system to assist clinicians to understand their patients' future vital signs to take appropriate actions against patients’ self-harm or physical violence against nursing staff or fellow patients.

\section{Framework}\label{framework}

In this section, the AI-enabled RPM framework proposed in this study will be discussed. This study was conducted in a simulated ward for real-time data collection. Each patient's vital signs such as heart rate, respiration, and physical activities will be extracted using passive RFID tags placed at different parts of the body as shown in Fig.~\ref{fig:framework}. To detect these passive RFID tags on the patient's body, two ultra-high frequency (UHF) 870 readers with integrated antennas were installed in the simulated ward. Four passive RFID tags were placed on different areas of the body such as the chest, abdomen, left arm, and right ankle of the patients. The tag in the chest area retrieves mechanical motions of heartbeat to estimate heart rate, and the tag at the abdomen extracts respiration rate based on contraction and expansion while breathing. With regard to physical activities, the two passive RFID tags placed at the right arm and left ankle retrieve limb movements of a patient.  The RFID tags were detected by the reader-antennas based on a measurement called received signal strength indicator (RSSI) which is the power received from the returned signal from a tag to the reader-antennas. As part of this data collection, the two UHF 870 antennas were fixed to the side walls of the simulated laboratory and the RSSI values retrieved from each passive tag were passed to the computer. Based on the RSSI values and frequency of the passive tag, the vital signs were processed, whereas physical activities were manually labelled using the phase orientation of the tags. Patient demographics such as height, weight, age, and gender were added to the extracted vital signs and labelled physical activities to enable personalised monitoring.
\begin{figure}
    \centering
    \includegraphics[scale=0.45]{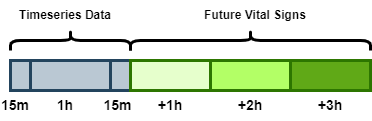}
    \caption{Prediction of Future Vital Signs}
    \label{fig:prediction}
\end{figure}

\subsection{Prediction Modelling}

In prediction modelling, the study is aimed to provide early detection of vital signs' deterioration based on time series prediction. The idea is to predict future vital signs of patients for the upcoming three hours. For this, the real-time streaming data being extracted from the RFID tags are segmented into windows of 1-hour size and the prediction can be provided every 15 minutes based on the previous 75 min (1-hour window size + 15 minutes) as shown in Fig.~\ref{fig:prediction}. In this prediction process, two vital signs heart rate and respiration for the upcoming 3 hours can be estimated based on the 75 minutes data samples.

The prediction study is formulated as a regression problem in which the extracted features from the segmented 1-hour windows of the vital signs. A multilayer perceptron (MLP)~\cite{taud2018multilayer} model which is a class of feed-forward artificial neural network model (ANN) is adopted for the regression modelling. The model consists of an input layer, three hidden layers, and an output layer with non-linearly activation nodes. Each node is a layer that connects to every node in the next layer with a certain weight $w$. Rectified linear unit (ReLU) activation function is used in each layer of the MLP regressor model. The activation function removes negative values by setting them to zero, as shown in Equation~\ref{eq:relu}. Considering the predicted value is a continuous numerical value, the loss function of the model is set to mean absolute error (MAE). 

\begin{equation}\label{eq:relu}
     {\displaystyle f(x)=\max(0,x)}
\end{equation}

The MLP regressor model prediction is mathematically presented as Equation~\ref{eq:mlpregressor} in which each input feature is multiplied with weight $w$ and added with a bias $b$. The updated features are passed through the activation function ReLU. Adam optimization algorithm runs averages of both the gradients and the second moments of the gradients. The Adam optimizer is used along with the MAE loss function in the model compilation.

\begin{equation}\label{eq:mlpregressor}
    {\displaystyle f(y)=\sum _{i=1}^{n} ReLU(b+w_{i}x_{i})}
\end{equation}
The vital signs processed from RFID passive tags data and patient demographics will be trained to the MLP model with an input node for each input variable in the input layer. The data will be further processed in three hidden layers with weight, bias, and activation units. The output layer will provide prediction results comprising future vital signs.

\subsection{Classification Modelling}
In classification modelling, 10 different activities such as standing still, climbing stairs, sitting and relaxing, lying down, walking, waist bends forward,  running, the frontal elevation of arms, knees bending, and  jump front \& back are classified based on the phase and orientation of RFID tags on a patient. A classification version of the MLP model, MLP Classifier, is used for the classification of physical activities. The classification model was configured with an input layer, three hidden layers, and an output layer. The hyper-parameters ReLU activation function and Adam optimiser in the classification model are similar to the prediction model, except for the loss function, which is changed to binary cross-entropy. The binary cross entropy has been calculated using the log loss function as shown in Equation~\ref{binarycrossentropy}. This loss function compares the predicted probabilities of the activity labels to the actual activity labels which result in a value ranging from 0 to 1 to observe how close the predicted probability of activity is close or far from the actual activity label. The classification problem in this study is multi-label, considering individual physical activity as a label with 1 to classify records of values. 

\begin{equation}\label{binarycrossentropy}
    {\displaystyle Log loss = \frac{1}{N} \sum _{i=1}^{N} -(y_{i}*log(p_{i})+(i-y_{i})*log(1-p_{i}))}
\end{equation}

\subsection{Evaluation}
In this study, two sets of evaluation metrics were used. The proposed framework can be evaluated with publicly available benchmark datasets like MIMIC-III~\cite{johnson2018mimic} and MHEALTH (Mobile HEALTH) dataset~\cite{banos2014mhealthdroid,nguyen2015recognizing}. The dataset can be processed for each subject for personalized monitoring. Individual subject data is split into 80\% training data for model learning and 20\% testing data for model evaluation. For the prediction model, the performance was evaluated using the metrics mean absolute error (MAE) and mean squared error (MSE). The predicted vital signs for input variables in testing data need to be compared to actual vital signs. The deviation of the predicted values from actual values can be measured in MAE and MSE. 

To evaluate the performance of the classification model, a traditional confusion matrix evaluation was adopted to estimate the precision, recall, and F1 score of each physical activity classification. A balanced accuracy metric was also adopted to estimate the model performance on individual physical activity classification. The trained classification model predicts the probability of each physical activity based on input variables in testing data. A threshold value needs to be set for the probability values of each record to classify the physical activities. After the classification, the model performance can be evaluated using the confusion matrix and the balanced accuracy.  

\section{Case Study}\label{casestudy}
The proposed AI framework can be adapted to mental healthcare facility settings for monitoring aggressive patients without touching their bodies. This can be achieved by installing UHF 870 RFID reader-antennas in a hospital ward, and arranging the passive RFID tags in a patient's hospital clothes at a readable range to the antennas. The RFID signals from the tags will need to be retrieved to a computer in a medical staff room via the reader-antennas for remote monitoring of the patient. The technological setup would assist medical staff in a mental health facility to monitor the patients who are aggressive or agitated and might cause harm to fellow patients, or staff or self-harm.  
\begin{figure*}[!h]
    \centering
    \includegraphics[scale=0.4]{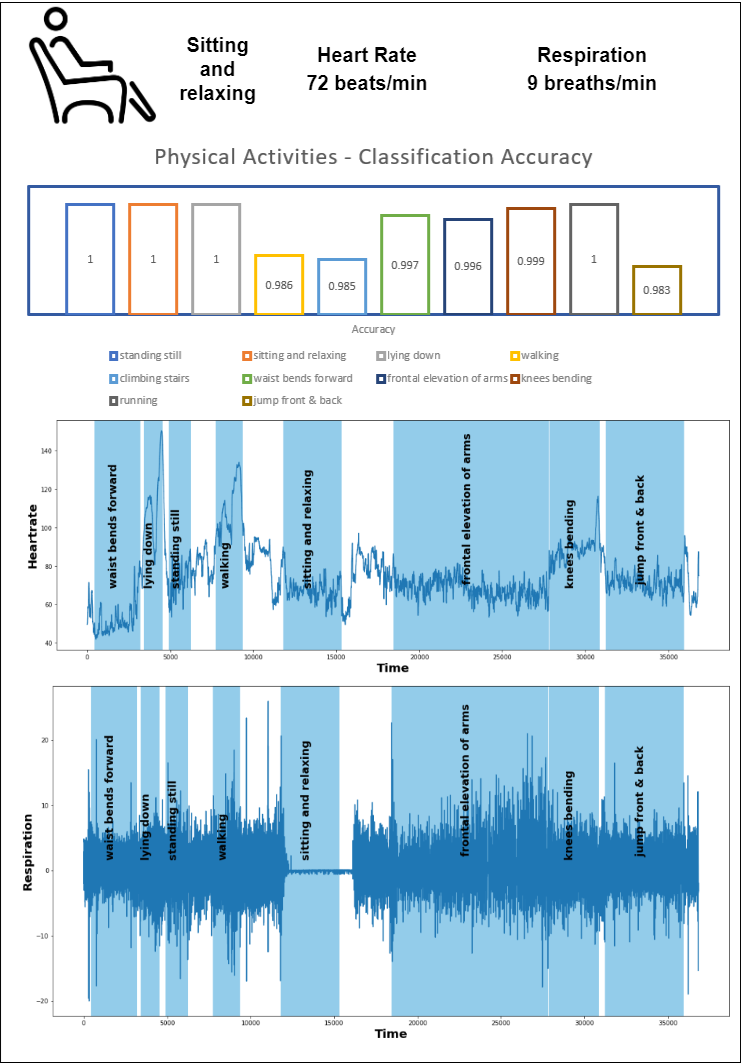}
    \caption{Output Plots}
    \label{fig:results}
\end{figure*}

\blockquote{David was a 47-year-old mentally challenged patient in a mental health facility in Queensland. He was suffering from post-traumatic stress disorder (PTSD) after sustaining a broken jaw, head injury and compound fracture of the left leg, in a road accident, 3 years earlier. He reported that in the accident, he had been driving a car and hit rail barriers. He had lost his closest friend on the spot and got trapped against the steering wheel and dashboard in an unconscious state. The patient left his job and joined the mental health facility for psychiatric care. He was suffering from PTSD symptoms like intrusive memories, and behaviour changes including quick temper, stress, sleep disturbance, social phobia, extreme panic attacks and anxiety~\cite{lethbridge2021australian}. Diagnosing and treating PTSD cannot be the same for any two patients as each patient experiences different symptoms. David had enrolled for inpatient treatments which include psychological interventions like acceptance commitment therapy and cognitive behavioural therapy. The patient's vital signs are fluctuating due to severe PTSD symptoms and had involved impulsive aggression due to failure of the prefrontal cortex to evaluate the impulse and weigh the consequences. The medical staff had experienced aggressive and agitated behaviour at times in the general ward. To prevent any self-harm and provide a safe environment to fellow patients and staff, precautionary measures were taken by moving David to a special ward in which the proposed AI-enabled RPM system is installed.}

In this case, the patient needs to be under continuous monitoring as there were severe symptoms of PTSD and this might lead to fluctuations in vital signs such as respiration and heart rate, deteriorating the patient's health status. Furthermore, the patient's behaviour was negatively related to anxiety, depression, and cause self-harm. This demands the need of monitoring the patient's physical activity. The patient was moved to a special ward with an RPM system installed and set up 4 passive RFID tags in his clothes to track his vital signs and external body movements. The patient was allowed to do his daily activities without any restrictions. The physical body movements and the vital signs were extracted from the passive RFID tags arranged in his hospital clothes. The retrieved parameters were combined with his demographics such as age, height, weight, body mass index (BMI), and so on. The data was then preprocessed to the train prediction and classification models to predict the patient's vital signs in the upcoming 3 hours, as well as classify his activities into the labelled 10 different activities. The patient's vital signs and physical actions were being monitored for 15 minutes using the data retrieved from RFID tags in the previous 75 minutes.

The output plots shown in Fig.~\ref{fig:results} present the current physical status of the patient, real-time heart rate and respiration readings. This is achieved based on real-time data collection from the tag data retrieved by the RFID reader antenna in the special hospital ward. In addition to these real-time readings, the classification model accuracy justifies and evident the physical status of the patient. The prediction model results are presented, in which heart rate and respiration for the upcoming 3 hours are forecasted. The forecasted vital signs are presented in alignment with the patient's physical activities. Line charts in Fig.~\ref{fig:results} are more sensible according to the patient's physical activities. The heart rate and respiration of the patient are fluctuations according to the physical activities.

\section{Conclusion and Future work}\label{conclusion}
Recent advancements in AI have contributed to many new techniques and increased the efficiency of healthcare applications. RPM systems are high in demand, and it requires efficient AI techniques to monitor patients in different health settings. This study presents the AI-enabled RPM system framework with the aim to monitor aggressive and agitated patients in a contained mental health facility. AI models were adopted to predict the vital signs of patients and classify their physical activities. The prediction model was able to predict future heart rate and respiration for the upcoming 3 hours based on the time-series data retrieved from passive RFID tags. The classification model can assist in classifying the current physical status of a patient by labelling to 10 physical activities. A case study of PTSD patients adopting the framework is illustrated in this study. This framework would assist medical staff to monitor their vulnerable patients with enhanced techniques of contact-less monitoring, can see the forecast of their health status, provide safety to patients and staff, and avoid unanticipated events like suicides.  

The proposed framework can be further enhanced to have an adaptive learning mechanism to learn the behaviour patterns and adapt this to the general ward where multiple patients are being treated while safeguarding their privacy using personalised monitoring techniques. Our future work will be focused on developing a deep reinforcement learning model~\cite{yu2021reinforcement}, considering each patient as an individual learning agent in a hospital environment trying to achieve maximum rewards by following the designed policy to achieve the goal to stay safe clinically. Large medical corpora including MIMIC~\cite{Johnson2016} database will be adapted to include a huge number of vital signs and train the reinforcement learning model.

\bibliographystyle{ACM-Reference-Format}
\bibliography{sample-base}

\end{document}